\documentclass[aps,prb,twocolumn,showpacs]{revtex4}

\usepackage{graphicx}
\usepackage{amsmath}
\usepackage{amssymb}

\begin{document}

\title{Doping dependence of charge order in electron-doped cuprate superconductors}

\author{Yingping Mou and Shiping Feng}
\email{spfeng@bnu.edu.cn}

\affiliation{Department of Physics, Beijing Normal University, Beijing 100875, China}

\begin{abstract}
In the recent studies of the unconventional physics in cuprate superconductors, one of the central issues is the interplay between charge order and superconductivity. Here the mechanism of the charge-order formation in the electron-doped cuprate superconductors is investigated based on the $t$-$J$ model. The experimentally observed momentum dependence of the electron quasiparticle scattering rate is qualitatively reproduced, where the scattering rate is highly anisotropic in momentum space, and is intriguingly related to the charge-order gap. Although the scattering strength appears to be weakest at the hot spots, the scattering in the antinodal region is stronger than that in the nodal region, which leads to the original electron Fermi surface is broken up into the Fermi pockets and their coexistence with the Fermi arcs located around the nodal region. In particular, this electron Fermi surface instability drives the charge-order correlation, with the charge-order wave vector that matches well with the wave vector connecting the hot spots, as the charge-order correlation in the hole-doped counterparts. However, in a striking contrast to the hole-doped case, the charge-order wave vector in the electron-doped side increases in magnitude with the electron doping. The theory also shows the existence of a quantitative link between the single-electron fermiology and the collective response of the electron density.
\end{abstract}


\maketitle

\section{Introduction}\label{Introduction}

The parent copper oxide compounds are half-filled Mott insulators, and then superconductivity emerges when holes or electrons are doped into these parent Mott insulators \cite{Bednorz86,Tokura89}. This Mott insulating-state appears to be due to the strong electron correlation \cite{Anderson87,Phillips10}, leading to that the multiple orders compete with superconductivity. In particular, the experimental data detected from various techniques \cite{Comin15,Campi15,Comin14,Wu11,Chang12,Ghiringhelli12,Neto14,Comin15a,Hashimoto15,Neto15,Neto16,Horio16,Armitage02,Matsui07,Armitage10} including scanning tunneling microscopy (STM), resonant X-ray scattering measurement (RXS), angle-resolved photoemission spectroscopy (ARPES), and other methods have demonstrated that the doping dependence of the charge-order correlation is a generic electronic property of the copper-oxide layers which is ubiquitous to all cuprate superconductors including both the hole- and electron-doped cuprate superconductors. In this case, the  microscopic origin of the charge-order formation has been hotly debated and believed to be key to the understanding of the problem of why cuprate superconductors have a high superconducting transition temperature.

The charge-order correlation is manifested itself by a periodic self-organization of the charge degrees of freedom in the system \cite{Comin15}. In the hole-doped cuprate superconductors, the combined STM, RXS measurement, and ARPES data showed that the charge-order correlation emergences consistently in surface and bulk, and in momentum and real space \cite{Comin15,Comin14,Wu11,Chang12,Ghiringhelli12,Neto14,Campi15,Comin15a,Hashimoto15}. In particular, this charge-order correlation is strongly doping dependent with the charge-order wave vector that smoothly {\it decreases} upon the increase of the hole doping \cite{Comin15,Comin14}. Moreover, the electron Fermi surface (EFS) measurements in the ARPES experiments revealed a close connection between the observed charge-order wave vector and the momentum vector connecting the tips of the Fermi arcs \cite{Comin15,Comin14,Neto14}, which in this case coincide with the hot spots on EFS, indicating that the hot spots on EFS play an important role in the formation of the charge-order correlation. This correspondence also shows the existence of a quantitative link between the collective response of the electron density and the single-electron fermiology \cite{Comin15,Comin14,Neto14}. On the electron-doped side, the magnetoresistance quantum oscillation measurements \cite{Helm09,Helm10,Kartsovnik11,Breznay16} and ARPES experimental observations \cite{Santander-Syro11} indicated the presence of a coexistence of the Fermi arcs and Fermi pockets around the nodal region. In particular, the very recent observations showed that the charge-order correlation with the wave vector that is consistent with the separation between straight segments of EFS occurs with similar periodicity, and along the same direction \cite{Neto15,Neto16,Horio16,Armitage02,Matsui07,Armitage10}, as the charge-order correlation in the hole-doped case \cite{Comin15,Comin14,Wu11,Chang12,Ghiringhelli12,Neto14,Campi15,Comin15a,Hashimoto15}. Furthermore, the charge-order wave vector $Q_{\rm CD}$ is reported to connect two hot spots \cite{Neto15,Neto16,Horio16,Armitage02,Matsui07,Armitage10}, and hence the electron quasiparticles are scattered between two hot spot regions connected by the charge-order wave vector and the same electron quasiparticle scattering causes the charge ordering instability. However, in a clear contrast to the hole-doped counterparts \cite{Comin15,Comin14}, the charge-order wave vector in the electron-doped side smoothly {\it increases} as a function of the electron doping \cite{Neto16}. These experimental results observed on both the hole- and electron-doped cuprate superconductors therefore raises questions \cite{Comin15,Campi15,Comin14,Wu11,Chang12,Ghiringhelli12,Neto14,Comin15a,Hashimoto15,Neto15,Neto16,Horio16,Armitage02,Matsui07,Armitage10}: (a) is there a common physical origin for the charge-order correlation in both the hole- and electron-doped cuprate superconductors? and (b) why there is a different doping dependence of the charge-order wave vector between the hole- and electron-doped cuprate superconductors?

Notably, the mechanism of the charge-order formation in the hole-doped cuprate superconductors has been recently taken forefront of the theoretical studies, where there is a general consensus that the charge-order correlation is driven by the EFS instability. Several attempts have been made to make these arguments more precise \cite{Lee14,Harrison14,Sachdev13,Meier13,Atkinson15,Feng16}. In our recent studies \cite{Feng16}, the physical origin of the charge-order correlation in the hole-doped cuprate superconductors has been discussed based on the $t$-$J$ model in the charge-spin separation fermion-spin representation, where we have shown that the charge-order state is driven by the EFS instability, with a characteristic wave vector corresponding to the hot spots on EFS. In particular, we have also shown that the charge-order wave vector smoothly {\it decreases} with the increases of the hole doping, in qualitative agreement with the experimental result \cite{Comin15,Comin14}. However, a comprehensive discussion of the electron-doped counterparts has not been given. In this paper, we try to study the mechanism of the charge-order formation in the electron-doped cuprate superconductors along with this line. We evaluate the electron self-energy in terms of the full charge-spin recombination and then employ it to calculate the electron spectral function and dynamical charge structure factor. In particular, we qualitatively reproduce the experimentally observed electron quasiparticle scattering rate along EFS \cite{Horio16}, where the electron quasiparticle scattering is highly anisotropic in momentum space, and opens a charge-order gap in the electron's band structure. The weakest scattering occurs exactly at the hot spots, however, the stronger scattering is found in the antinodal region than in the nodal region. This special structure of the electron quasiparticle scattering rate (then the charge-order gap) on EFS therefore leads to that only part of EFS survives as the disconnected segments around the nodal region, and then the tips of these disconnected segments assemble on the hot spots to form the Fermi pockets, generating a coexistence of the Fermi arcs and Fermi pockets. Moreover, this EFS instability drives the charge-order correlation, with the charge-order wave vector that matches well with the wave vector connecting the hot spots on EFS, in a striking similarity with the charge-order wave vector in the hole-doped counterparts, indicating a common physical origin for the charge-order correlation in both the hole- and electron-doped cuprate superconductors. However, in a clear contrast to the hole-doped case, the charge-order wave vector in the electron-doped cuprate superconductors smoothly {\it increases} with the increase of the electron doping.

The rest of this paper is organized as follows. The basic formalism is presented in Sec. \ref{Formalism}, where we generalize the calculation of the single-electron Green's function and the related electron spectrum function in the hole-doped case in Ref. \cite{Feng16} to the present case in the electron-doped side. Based on this electron spectral function, the doping dependence of the charge-order correlation in the electron-doped cuprate superconductors is discussed in Sec. \ref{characteristics}, where we evaluate the dynamical charge structure factor in terms of the electron spectral function, and then show that an resonance peak appears with the characteristic wave vector that is well consistent with the wave vector connected by the hot spots. Moreover, the suppression of the peak by tuning the energy away from the resonance confirms the presence of the charge-order correlation in the electron-doped cuprate superconductors. Our results therefore also shows the existence of a quantitative link between the electron quasiparticle excitations determined by the electronic structure and the charge-order correlation determined by the collective response of the electron density. Finally, we give a summary and discussions in Sec. \ref{conclusions}.

\section{Formalism}\label{Formalism}

To elucidate the strong electron correlation and the resulting charge ordering instability in the electron-doped cuprate superconductors, the exact knowledge of EFS and its evolution with the electron doping is of crucial importance. EFS is determined by the poles of the single-electron Green's function,
\begin{eqnarray}\label{EGF}
G({\bf k},\omega)={1\over\omega-\varepsilon_{\bf k}-\Sigma_{1}({\bf k},\omega)},
\end{eqnarray}
where $\varepsilon_{\bf k}$ gives the dispersion of the bare electron excitation spectrum, while $\Sigma_{1}({\bf k},\omega)$ is the electron self-energy, which alone captures the essential physics of the strong electron correlation, and therefore dominates the detailed properties of the electronic state \cite{Armitage10,Damascelli03,Campuzano04}. The electron spectral function is directly related to the imaginary part of the single-electron Green's function (\ref{EGF}), and can be expressed explicitly as,
\begin{eqnarray}\label{SF}
A({\bf k},\omega)&=& -2{\rm Im}G({\bf k},\omega)\nonumber\\
&=&{2|{\rm Im}\Sigma_{1}({\bf k},\omega)|\over [\omega-\varepsilon_{\bf k}-{\rm Re}\Sigma_{1}({\bf k}, \omega)]^{2}+[{\rm Im}\Sigma_{1}({\bf k}, \omega)]^{2}},~~~~~
\end{eqnarray}
then the electron spectrum and the related electronic state properties can be measured in ARPES experiments \cite{Damascelli03,Campuzano04,Armitage10}, where ${\rm Im}\Sigma_{1}({\bf k},\omega)$ and ${\rm Re}\Sigma_{1}({\bf k}, \omega)$ are, respectively, the corresponding imaginary and real parts of the electron self-energy $\Sigma_{1}({\bf k},\omega)$. In particular, this imaginary part ${\rm Im} \Sigma_{1}({\bf k},\omega)$ is also identified as the electron quasiparticle dynamical scattering rate $\Gamma({\bf k},\omega)$,
\begin{eqnarray}\label{DSR}
\Gamma({\bf k},\omega)={\rm Im} \Sigma_{1}({\bf k},\omega).
\end{eqnarray}

It is widely accepted by now the simplest model which contains in itself the interacting spin and charge degrees of freedom is so-called $t$-$J$ model on a square-lattice \cite{Anderson87,Phillips10},
\begin{eqnarray}\label{tjham}
H=\sum_{\langle l\hat{a}\rangle\sigma}t_{l\hat{a}}C^{\dagger}_{l\sigma}C_{l+\hat{a}\sigma}+\mu\sum_{l\sigma}C^{\dagger}_{l\sigma}C_{l\sigma} +J\sum_{\langle l \hat{\eta}\rangle}{\bf S}_{l}\cdot {\bf S}_{l+\hat{\eta}},~~
\end{eqnarray}
where we consider the case with the hopping integrals that are $t_{l\hat{a}}=t$ for the nearest-neighbor (NN) sites $\hat{a}=\hat{\eta}$, $t_{l\hat{a}}=-t'$ for the next NN sites $\hat{a}=\hat{\tau}$, and $t_{l\hat{a}}=0$ otherwise, with $t<0$ and $t'<0$ in the electron-doped side, $C_{l\sigma}^{\dagger}$ and $C_{l\sigma}$ are the electron operators that respectively create and annihilate electrons with spin $\sigma$ on the site $l$, ${\bf S}_{l}$ is a local spin operator, and $\mu$ is the chemical potential, while $\langle l\hat{a}\rangle$ means that $l$ runs over all sites, and for each $l$, over its NN sites $\hat{a}=\hat{\eta}$ or the next NN sites $\hat{a}=\hat{\tau}$. Although the values of the parameters $J$, $t$, and $t'$ are believed to vary somewhat from compound to compound \cite{Armitage10,Hozoi08}, however, as a qualitative discussion in this paper, the parameters are chosen as \cite{Armitage10,Hozoi08} $t/J=-2.5$ and $t'/t=0.4$. In the electron-doped side, this $t$-$J$ model (\ref{tjham}) is imposed a on-site local constraint $\sum_{\sigma}C^{\dagger}_{l\sigma}C_{l\sigma}\geq 1$ in order to ensure no-zero electron occupancy of any lattice site. Since the strong electron correlation in the $t$-$J$ model (\ref{tjham}) manifests itself by this no-zero electron occupancy local constraint, and therefore it is crucially important to treat this local constraint properly. Apart from the numerical approaches \cite{Dagotto94}, a popular analytical strategy to enforce the electron local constraint is the approach based on the charge-spin separation \cite{Yu92,Feng93,Anderson00,Lee06,Feng9404,Feng15}. In particular, a charge-spin separation fermion-spin approach has been developed for a proper treatment of no-double electron occupancy local constraint in the hole-doped case \cite{Feng9404,Feng15}. To apply this fermion-spin approach to the electron-doped side, we can work in the hole representation via a particle-hole transformation $C_{l\sigma}\rightarrow f^{\dagger}_{l-\sigma}$, and then the $t$-$J$ model (\ref{tjham}) can be rewritten in the hole representation as,
\begin{eqnarray}\label{tjham1}
H=-\sum_{\langle l\hat{a}\rangle\sigma}t_{l\hat{a}}f^{\dagger}_{l+\hat{a}\sigma}f_{l\sigma}-\mu\sum_{l\sigma}f^{\dagger}_{l\sigma}f_{l\sigma} +J\sum_{\langle l \hat{\eta}\rangle}{\bf S}_{l}\cdot {\bf S}_{l+\hat{\eta}},~~~
\end{eqnarray}
where $f^{\dagger}_{l\sigma}$ ($f_{l\sigma}$) is the hole creation (annihilation) operator, and then the local constraint of no-zero electron occupancy $\sum_{\sigma}C^{\dagger}_{l\sigma}C_{l\sigma}\geq 1$ in the electron representation is therefore transferred as the local constraint of no-double occupancy $\sum_{\sigma} f^{\dagger}_{l\sigma} f_{l\sigma}\leq 1$ in the hole representation. This local constraint of no-double occupancy now can be dealt properly within the framework of the fermion-spin theory \cite{Feng9404,Feng15}, $f_{l\uparrow}=a^{\dagger}_{l\uparrow} S^{-}_{l}$ and $f_{l\downarrow}=a^{\dagger}_{l\downarrow}S^{+}_{l}$, where the spinful fermion operator $a_{l\sigma}=e^{-i\Phi_{l\sigma}}a_{l}$ carries the charge of the constrained hole together with some effects of spin configuration rearrangements due to the presence of the doped electron itself (charge carrier), while the spin operator $S_{l}$ represents the spin degree of freedom of the constrained hole, and then the local constraint of no-double occupancy is satisfied in the actual calculations. In this fermion-spin approach, the scattering from spins due to the  charge-carrier fluctuations dominates the spin response, while the charge-spin recombination of a charge carrier and a localized spin automatically gives the electron quasiparticle character.

In the framework of the charge-spin separation fermion-spin theory, the single-hole Green's function $G_{\rm f}({\bf k},\omega)$ of the $t$-$J$ model (\ref{tjham1}) in the hole representation on the other hand can be obtained in terms of the charge-spin recombination. Recently, we \cite{Feng15a} have developed a full charge-spin recombination scheme to fully recombine a charge carrier and a localized spin into a constrained electron, where the obtained electron propagator can give a consistent description of EFS in the hole-doped cuprate superconductors \cite{Feng16}. However, a question is how we can obtain explicitly the single-electron Green's function (\ref{EGF}) based on the $t$-$J$ model (\ref{tjham}). In the following discussions, we give only the main details in the calculations of the single-electron Green's function of the $t$-$J$ model. Following the previous discussions \cite{Feng16,Feng15a}, the single-hole Green's function of the $t$-$J$ model (\ref{tjham1}) can be obtained in terms of the full charge-spin recombination scheme as,
\begin{eqnarray}\label{HGF}
G_{\rm f}({\bf k},\omega)={1\over\omega-\varepsilon^{({\rm f})}_{\bf k}-\Sigma^{({\rm f})}_{1}({\bf k},\omega)},
\end{eqnarray}
where $\varepsilon^{({\rm f})}_{\bf k}=-Zt\gamma_{\bf k}+Zt'\gamma_{\bf k}'+\mu$ is the bare hole excitation spectrum, with $\gamma_{\bf k}=({\rm cos}k_{x}+{\rm cos}k_{y})/2$, $\gamma_{\bf k}'={\rm cos}k_{x}{\rm cos}k_{y}$, and the number of the NN or next NN sites on a square lattice $Z$, while the hole self-energy $\Sigma^{({\rm f})}_{1}({\bf k},\omega)$ due to the interaction between holes by the exchange of spin excitations can be evaluated explicitly in terms of the spin bubble as \cite{Feng16,Feng15a},
\begin{eqnarray}\label{ESE}
\Sigma^{({\rm f})}_{1}({\bf k},\omega)&=&{1\over N^{2}}\sum_{{\bf pp'}\nu=1,2}(-1)^{\nu+1}\Omega^{({\rm f})}_{\bf pp'k}\left ({F^{(\nu)}_{1{\rm f} {\bf pp'k}}\over \omega+\omega_{\nu{\bf p}{\bf p}'}-\bar{\varepsilon}^{({\rm f})}_{{\bf p}+{\bf k}}} \right. \nonumber\\
&+&\left . {F^{(\nu)}_{2{\rm f}{\bf pp'k}}\over\omega-\omega_{\nu{\bf p}{\bf p}'}-\bar{\varepsilon}^{({\rm f})}_{{\bf p}+{\bf k}}}\right ),~~~~~~~
\end{eqnarray}
where $\Omega^{({\rm f})}_{\bf pp'k}=Z^{({\rm f})}_{\rm F}\Lambda^{2}_{{\bf p}+{\bf p}'+{\bf k}}B_{{\bf p}'}B_{{\bf p}+{\bf p}'}/ (4\omega_{{\bf p}'}\omega_{{\bf p}+{\bf p}'})$, $\Lambda_{\bf k}=Zt\gamma_{\bf k}-Zt'\gamma_{\bf k}'$, $\omega_{\nu{\bf p}{\bf p}'}=\omega_{{\bf p} +{\bf p}'}-(-1)^{\nu}\omega_{\bf p'}$, $\bar{\varepsilon}^{({\rm f})}_{\bf k}=Z^{({\rm f})}_{\rm F}\varepsilon^{({\rm f})}_{\bf k}$, the single-particle coherent weight $Z^{({\rm f})-1}_{\rm F}=1-\Sigma^{({\rm f})}_{1{\rm o}}({\bf k},\omega=0)|_{{\bf k}=[\pi,0]}$, with $\Sigma^{({\rm f})}_{1{\rm o}}({\bf k}, \omega)$ that is the antisymmetric part of $\Sigma^{({\rm f})}_{1}({\bf k},\omega)$, the functions $F^{(\nu)}_{1{\rm f}{\bf pp'k}}=n_{\rm F} (\bar{\varepsilon}^{({\rm f})}_{{\bf p}+{\bf k}})n^{(\nu)}_{{\rm 1B}{\bf pp'}}+n^{(\nu)}_{{\rm 2B}{\bf pp'}}$, $F^{(\nu)}_{2{\rm f}{\bf pp'k}}=[1- n_{\rm F}(\bar{\varepsilon}^{({\rm f})}_{{\bf p}+{\bf k}})]n^{(\nu)}_{{\rm 1B}{\bf pp'}}+n^{(\nu)}_{{\rm 2B}{\bf pp'}}$, with $n^{(\nu)}_{{\rm 1B} {\bf pp'}}=1+n_{\rm B}(\omega_{{\bf p}'+{\bf p}})+n_{\rm B}[(-1)^{\nu+1}\omega_{\bf p'}]$, $n^{(\nu)}_{{\rm 2B}{\bf pp'}}=n_{\rm B}(\omega_{{\bf p}' +{\bf p}})n_{\rm B}[(-1)^{\nu+1}\omega_{\bf p'}]$, and $n_{\rm B}(\omega)$ and $n_{\rm F}(\omega)$ that are the boson and fermion distribution functions, respectively. The expression form of the mean-field spin excitation spectrum $\omega_{\bf k}$, and function $B_{\bf k}$ have been given explicitly in Ref. \onlinecite{Cheng08}, while the single-particle coherent weight $Z^{({\rm f})}_{\rm F}$ together with the chemical potential can be determined self-consistently.

We now turn to evaluate the single-electron Green's function $G({\bf k},\omega)$ in Eq. (\ref{EGF}) based on the $t$-$J$ model (\ref{tjham}), which is directly related to the single-hole Green's function $G_{\rm f}({\bf k},\omega)$ in Eq. (\ref{HGF}) in terms of the particle-hole transformation as $G(l-l',t-t')= \langle\langle C_{l\sigma}(t); C^{\dagger}_{l'\sigma}(t')\rangle\rangle=\langle\langle f^{\dagger}_{l\sigma}(t);f_{l'\sigma}(t') \rangle \rangle=-G_{\rm f} (l'-l,t'-t)$. With the help of the above single-hole Green's function (\ref{HGF}), the single-electron Green's function (\ref{EGF}) of the $t$-$J$ model (\ref{tjham}) is therefore obtained as $G({\bf k},\omega)=-G_{\rm f}({\bf k},-\omega)$, with the bare electron excitation spectrum and electron self-energy that are obtained as $\varepsilon_{\bf k}=-\varepsilon^{({\rm f})}_{\bf k}$ and $\Sigma_{1}({\bf k}, \omega)=- \Sigma^{({\rm f})}_{1}({\bf k},-\omega)$, respectively. In this case, the electron quasiparticle dynamical scattering rate $\Gamma({\bf k}, \omega)$ in Eq. (\ref{DSR}) arises from the electron self-energy due to the interaction between electrons by the exchange of spin excitations.

\begin{figure*}[t!]
\centering
\includegraphics[scale=1.0]{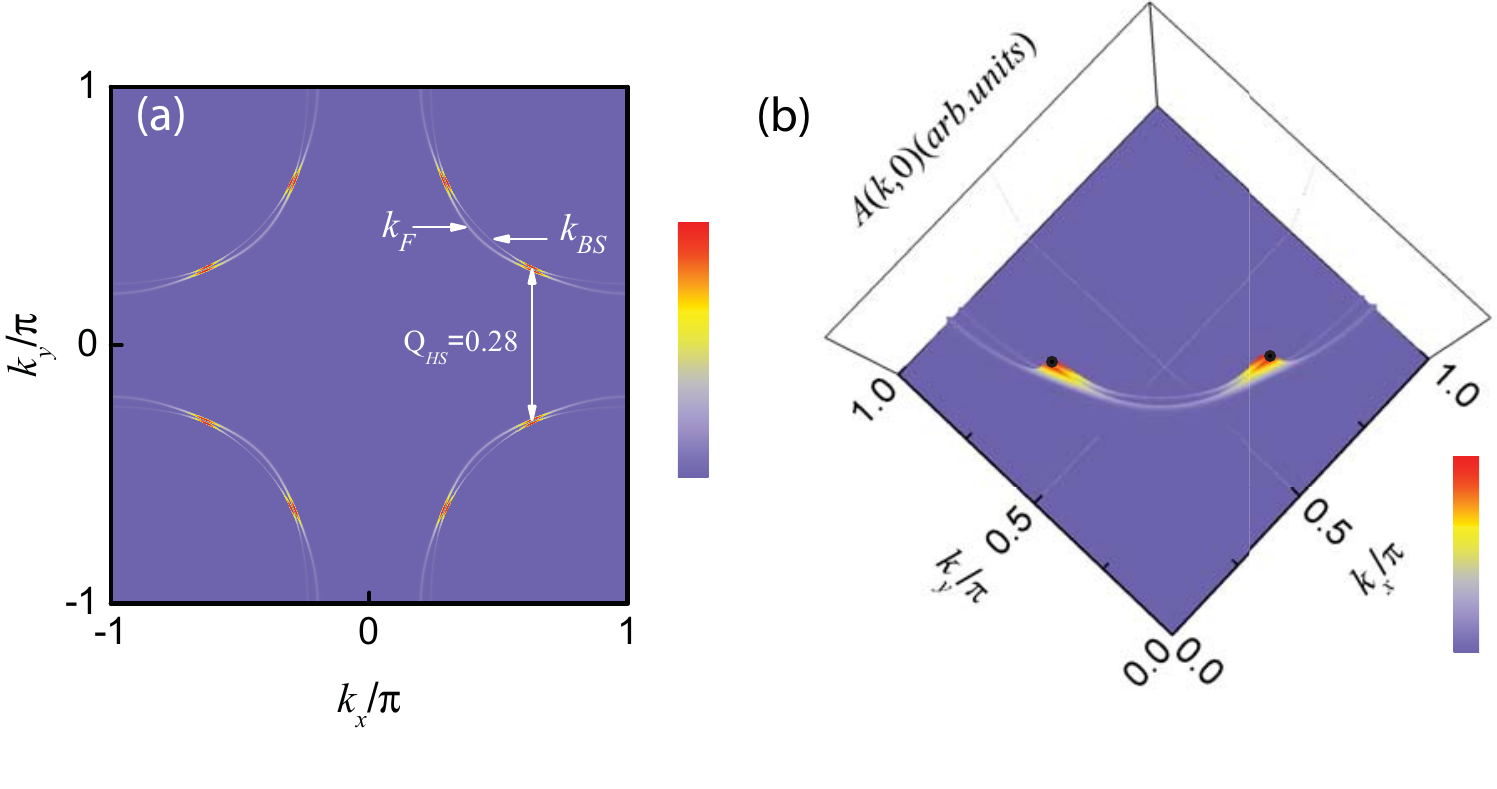}
\caption{(Color online) (a) The map of the electron spectral intensity $A({\bf k},0)$ and (b) the electron spectral function $A({\bf k},0)$ in the $[k_{x},k_{y}]$ plane at $\delta=0.15$ with $T=0.002J$ for $t/J=-2.5$ and $t'/t=0.4$.
\label{spectral-maps}}
\end{figure*}

\section{Doping dependence of charge order}\label{characteristics}

In this section, we discuss the doping dependence of the charge-order correlation, and show that the strong electron correlation in the electron-doped cuprate superconductors induces an EFS instability, which then drives the charge-order correlation. In particular, we also show that there is indeed a quantitative link between the electron quasiparticle excitations determined by the low-energy electronic structure and the charge-order correlation determined by the collective response of the electron density.

\subsection{Electron Fermi surface reconstruction}\label{Fermi-arcs}

The charge-order correlation is defined as an electronic phase breaking translational symmetry via a self-organization of the electrons into periodic structures incompatible with the periodicity of the underlying lattice \cite{Comin15}, such an aspect should be reflected in the electronic structure. In Fig. \ref{spectral-maps}, we plot (a) the map of the electron spectral intensity $A({\bf k},0)$ in Eq. (\ref{SF}) and (b) the electron spectral function $A({\bf k}, 0)$ in the $[k_{x},k_{y}]$ plane at the electron doping $\delta=0.15$ with temperature $T=0.002J$ for parameters $t/J=-2.5$ and $t'/t=0.4$. The hot spots on EFS are marked by the black circles in Fig. \ref{spectral-maps}b, where the spectral intensity exhibits a largest value. Our main results in Fig. \ref{spectral-maps} are summarized as: (A) the single continuous contour in momentum space with a uniform distribution of the low-energy spectral weight of the electron quasiparticle excitations in the case of the absence of the strong electron correlation has been split by the electron self-energy $\Sigma_{1}({\bf k},0)$ [then the strong electron correlation] into two contours ${\bf k}_{\rm F}$ and ${\bf k}_{\rm BS}$, respectively. In particular, this low-energy spectral weight redistribution leads to an EFS reconstruction; (B) the electron spectral intensity at the contours ${\bf k}_{\rm F}$ and ${\bf k}_{\rm BS}$ are suppressed by the electron self-energy, however, this suppression is highly anisotropic, where the electron spectral intensity around the antinodal region is suppressed heavily, while the electron self-energy has a more modest effect on the electron spectral intensity around the nodal region, which leads to that the low-energy electron quasiparticles mainly reside at the disconnected segments around the nodal region; (C) however, the tips of the disconnected segments on the contours ${\bf k}_{\rm F}$ and ${\bf k}_{\rm BS}$ assemble on the hot spots to form a Fermi pocket, generating a coexistence of the Fermi arcs and Fermi pockets, with the disconnected segment around the nodal region at the contour ${\bf k}_{\rm F}$ that is so-called the Fermi arc, and is also defined as the front side of the Fermi pocket, while the other at the contour ${\bf k}_{\rm BS}$ around the nodal region is associated with the back side of the Fermi pocket. These results are similar to these observed in the hole-doped case \cite{Yang08,Chang08,Meng09,Yang11,Zhao17}, and also are in qualitative agreement with the experimental data obtained by means of the magnetoresistance quantum oscillation \cite{Helm09,Helm10,Kartsovnik11,Breznay16} and ARPES experimental measurements \cite{Santander-Syro11}, where the well defined Fermi pockets and their coexistence with the Fermi arcs are observed around the nodal region. In particular, it should be noted that the very recent ARPES observations show that the EFS reconstruction and related low-energy electron structure in the electron-doped cuprate superconductors is closely related with the annealing effect \cite{Horio16,Song17}, and then the annealing and oxygen vacancy induce a sufficient change in the charge carrier density. Moreover, upon the proper annealing, the EFS reconstruction and related low-energy electron structure are very similar to these observed in the hole-doped counterparts, providing the experimental evidences of the absence of the disparity between the phase diagram of the hole- and electron-doped cuprate superconductors. The present results are also consistent with these recently experimental data \cite{Horio16,Song17}.

\subsection{Charge-order correlation driven by electron Fermi surface instability}\label{charge-order}

The results in Fig. \ref{spectral-maps} also show that the partial spectral weight at the Fermi arc has been pushed to the back side of the Fermi pocket due to the spectral weight redistribution in the presence of the strong electron correlation, which leads to that although both the Fermi arc and back side of the Fermi pocket possess finite spectral weight, the electron quasiparticle peaks are anomalously broad at both the Fermi arc and back side of the Fermi pocket, while the very sharp electron quasiparticle peaks appear at the hot spots, where the most of the electron quasiparticles are accommodated. This is why the electron quasiparticle excitations around the hot spots contribute effectively to the electron quasiparticle scattering process \cite{Horio16}. In particular, the electron quasiparticle scattering wave vector between the hot spots on the straight Fermi arcs in Fig. \ref{spectral-maps}a at the electron doping $\delta\sim 0.15$ is found to be $Q_{\rm HS}\sim 0.28$ (hereafter we use the reciprocal units), which is well consistent with experimental average value \cite{Neto15,Neto16,Horio16,Armitage02,Matsui07} of the charge-order wave vector $Q_{\rm CD}\sim 0.27$ oberved in the electron-doped cuprate superconductors around the electron doping $\delta\sim 0.15$, indicating that the charge-order correlation in the electron-doped cuprate superconductors is driven by the EFS instability. In particular, this behavior resembles the observations for the hole-doped case \cite{Comin15,Comin14,Neto14,Feng16}, and therefore there is a common physical origin for the charge-order correlation in both the hole- and electron-doped cuprate superconductors.

\begin{figure}[h!]
\centering
\includegraphics[scale=1.2]{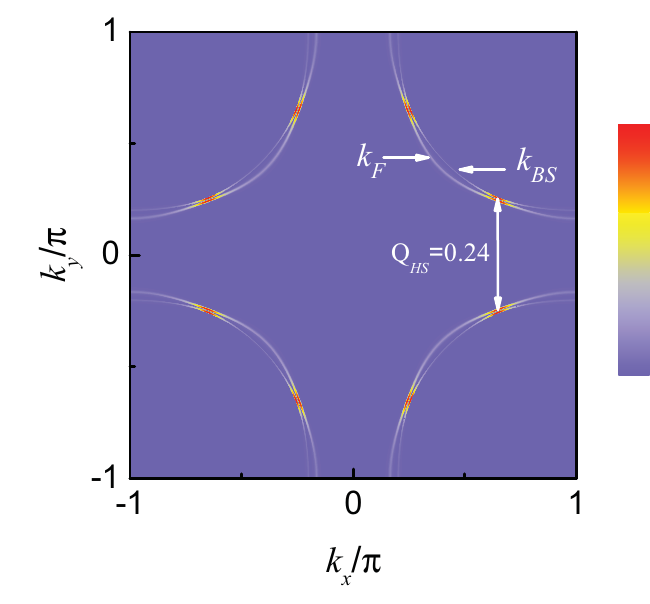}
\caption{(Color online) The map of the electron spectral intensity $A({\bf k},0)$ at $\delta=0.09$ with $T=0.002J$ for $t/J=-2.5$ and $t'/t=0.4$.
\label{spectral-maps-doping}}
\end{figure}

\begin{figure}[h!]
\centering
\includegraphics[scale=1.1]{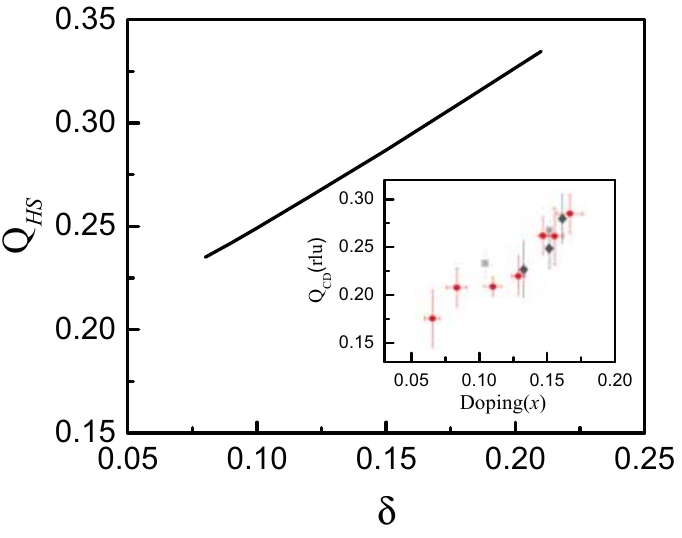}
\caption{(Color online) The charge-order wave vector as a function of doping for $t/J=-2.5$ and $t'/t=0.4$ with $T=0.002J$. Inset: the corresponding experimental data of the electron-doped cuprate superconductors taken from Ref. \onlinecite{Neto16}. \label{CD-doping}}
\end{figure}

As a natural consequence of the doped Mott insulators, this charge-order correlation is also doping dependent. In Fig. \ref{spectral-maps-doping}, we plot the map of the electron spectral intensity $A({\bf k},0)$ at the electron doping $\delta=0.09$ with $T=0.002J$ for $t/J=-2.5$ and $t'/t=0.4$. Comparing it with Fig. \ref{spectral-maps}a for the same set of parameters except for the electron doping $\delta=0.09$, we therefore find that with the {\it decrease} of the electron doping, the hot spots move away from the nodal points, in a clear contrast to the hole-doped case, where the hot spots move towards to the nodal points when the hole doping is decreased \cite{Comin14,Feng16}. This unusually doping dependent hot-spot positions in the electron-doped side therefore leads to that the charge-order wave vector increases as a function of the electron doping. To see this point more clearly, we have made a series of calculations for the electron spectral function $A({\bf k},0)$ at different electron doping concentrations, and the result for the extracted wave vector $Q_{\rm HS}$ as a function of the electron doping with $T=0.002J$ for $t/J=-2.5$ and $t'/t=0.4$ is plotted in Fig. \ref{CD-doping} in comparison with the corresponding experimental data \cite{Neto16} of the charge-order wave vector $Q_{\rm CO}$ observed on the electron-doped cuprate superconductors, where $Q_{\rm HS}$ {\it increases} linearly with the electron doping, also in qualitative agreement with the experimental data \cite{Neto16}.

In the case of the absence of the strong electron correlation, the overall shape of EFS obtained from the map of the electron spectral function at zero energy is a continuous contour centered at $[\pi,\pi]$ of the Brillouin zone, where the electron quasiparticle excitation spectrum is gapless, and then the electron quasiparticle lifetime on EFS is infinitely long. However, when the strong electron correlation is contained by the electron self-energy $\Sigma_{1}({\bf k},\omega)$, the electron quasiparticle energies are heavily renormalized by ${\rm Re}\Sigma_{1}({\bf k},\omega)$ and they acquire a finite lifetime $\tau({\bf k},\omega)=\Gamma^{-1}({\bf k},\omega)$. In this case, the essential physics of the EFS reconstruction and the related EFS instability driven charge order are naturally associated with the strong electron correlation. This states a fact that the locations of the continuous contours in momentum space are determined directly by the poles of the electron Green's function (\ref{EGF}) at zero energy, while the intensity of the low-energy electron excitation spectrum at the continuous contours is inversely proportional to the electron quasiparticle scattering rate $\Gamma({\bf k},\omega)$. However, as we have shown in the previous discussions for the hole-doped case \cite{Feng16,Zhao17}, the electron self-energy $\Sigma_{1}({\bf k},\omega)$ in Eq. (\ref{ESE}) can be also rewritten as,
\begin{eqnarray}\label{PG}
\Sigma_{1}({\bf k},\omega)\approx {[\bar{\Delta}_{\rm CD}({\bf k})]^{2}\over\omega+\varepsilon_{0{\bf k}}},
\end{eqnarray}
where the corresponding energy spectrum $\varepsilon_{0{\bf k}}$ and the momentum dependence of the charge-order gap $\bar{\Delta}_{\rm CD}({\bf k})$ can be obtained directly from the electron self-energy $\Sigma_{1}({\bf k},\omega)$ in Eq. (\ref{ESE}) and its antisymmetric part $\Sigma_{\rm 1o} ({\bf k},\omega)$ as $\varepsilon_{0{\bf k}}=-\Sigma_{1}({\bf k},0)/\Sigma_{\rm 1o}({\bf k},0)$ and $\bar{\Delta}_{\rm CD}({\bf k})=\Sigma_{1}({\bf k},0)/\sqrt{-\Sigma_{\rm 1o}({\bf k},0)}$, respectively. In particular, in analogy to the pseudogap effect in the hole-doped case \cite{Feng16,Zhao17}, this charge-order gap $\bar{\Delta}_{\rm CD}({\bf k})$ in the electron-doped side can be also identified as being a region of the electron self-energy effect in which the charge-order gap $\bar{\Delta}_{\rm CD}({\bf k})$ suppresses the electron spectral intensity. Moreover, the dynamical electron quasiparticle scattering rate $\Gamma({\bf k},\omega)$ can be expressed explicitly in terms of the charge-order gap $\bar{\Delta}_{\rm CD}({\bf k})$ as,
\begin{eqnarray}\label{IESE}
\Gamma({\bf k},\omega)={\rm Im}\Sigma_{1}({\bf k},\omega)\approx  2\pi[\bar{\Delta}_{\rm CD}({\bf k})]^{2}\delta(\omega+\varepsilon_{0{\bf k}}),
\end{eqnarray}
which therefore can be obtained from the width of the electron quasiparticle peaks in the ARPES experiments \cite{Horio16}, and is intimately related to the charge-order gap $\bar{\Delta}_{\rm CD}({\bf k})$ in the electron-doped cuprate superconductors. On the other hand, Eq. (\ref{IESE}) also shows that the product of the charge-order gap $[\bar{\Delta}_{\rm CD}({\bf k})]^{2}$ and the delta function $\delta(\omega+\varepsilon_{0{\bf k}})$ has the same momentum dependence and plays the same role in the suppression of the electron spectral weight as that of the electron quasiparticle scattering rate $\Gamma({\bf k},\omega)$. According to the electron self-energy $\Sigma_{1}({\bf k},\omega)$ in Eq. (\ref{PG}), the single-electron Green's function in Eq. (\ref{EGF}) can be rewritten as,
\begin{eqnarray}\label{EGF1}
G({\bf k},\omega)\approx {W^{+}_{\bf k}\over\omega-E^{+}_{\bf k}-i\Gamma({\bf k},\omega)}+{W^{-}_{\bf k}\over\omega-E^{-}_{\bf k}-i\Gamma({\bf k}, \omega)}, ~~~
\end{eqnarray}
where $W^{+}_{\bf k}=(E^{+}_{\bf k}+\varepsilon_{0{\bf k}})/(E^{+}_{\bf k}-E^{-}_{\bf k})$ and $W^{-}_{\bf k}=-(E^{-}_{\bf k}+\varepsilon_{0{\bf k}})/(E^{+}_{\bf k}-E^{-}_{\bf k})$ are the coherence factors with the constraint $W^{+}_{\bf k}+W^{-}_{\bf k}=1$ for any wave vector ${\bf k}$ (the sum rule). As a natural consequence of the presence of the charge-order gap, the electron energy band has been split into the antibonding band $E^{+}_{\bf k}=[\varepsilon_{\bf k}-\varepsilon_{0{\bf k}}+\sqrt{(\varepsilon_{\bf k}+\varepsilon_{0{\bf k}})^{2}+4\bar{\Delta}^{2}_{\rm CD}({\bf k}) }]/2$ and bonding band $E^{-}_{\bf k} =[\varepsilon_{\bf k}-\varepsilon_{0{\bf k}}-\sqrt{(\varepsilon_{\bf k}+\varepsilon_{0{\bf k}})^{2}+ 4\bar{\Delta}^{2}_{\rm CD}({\bf k})} ]/2$, respectively. This band splitting induced by the charge-order gap therefore leads to form two contours ${\bf k}_{\rm F}$ and ${\bf k}_{\rm BS}$ in momentum space shown in Fig. \ref{spectral-maps}.

\begin{figure*}[t!]
\centering
\includegraphics[scale=1.0]{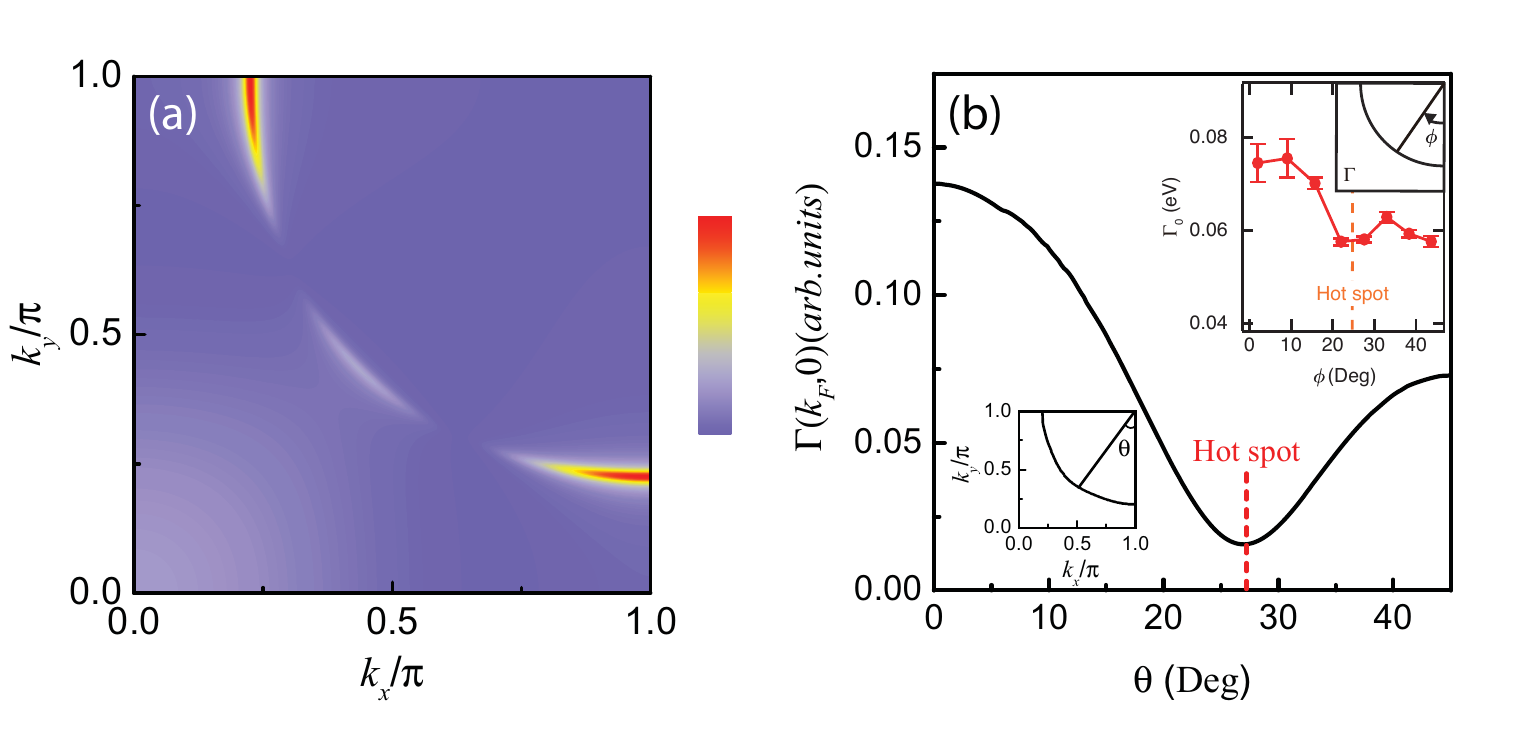}
\caption{(Color online) (a) The map of the electron quasiparticle scattering rate and (b) the angular dependence of the electron quasiparticle scattering rate along ${\bf k}_{\rm F}$ from the antinode to the node at $\delta=0.15$ with $T=0.002J$ for $t/J=-2.5$ and $t'/t=0.4$. The position of the hot spot in (b) is indicated by the dashed vertical line. Inset in (b): the corresponding experimental data of Pr$_{1.3-x}$La$_{0.7}$Ce$_{x}$CuO$_{4}$ taken from Ref. \onlinecite{Horio16}. \label{scattering-rate}}
\end{figure*}

From the above expression of the single-electron Green's function (\ref{EGF1}), now we find that in analogy to the hole-doped case \cite{Zhao17}, the first contour ${\bf k}_{\rm F}$ in Fig. \ref{spectral-maps}a is determined by the poles of the first term of the right-hand side of the single-electron Green's function (\ref{EGF1}) at zero energy, where the electron antibonding dispersion $E^{+}_{\bf k}$ along ${\bf k}_{\rm F}$ is equal to zero, while the second contour ${\bf k}_{\rm BS}$ in Fig. \ref{spectral-maps}a is determined by the poles of the second term of the right-hand side of the single-electron Green's function (\ref{EGF1}) at zero energy, where the electron bonding dispersion $E^{-}_{\bf k}$ along ${\bf k}_{\rm BS}$ is equal to zero. Since the electron self-energy $\Sigma_{1}({\bf k},\omega)$ originates in the electron's coupling to spin excitations, the charge-order gap $\bar{\Delta}_{\rm CD}({\bf k})$ is strong dependence of momentum. To reveal this highly anisotropic $\bar{\Delta}_{\rm CD}({\bf k})$ in momentum space clearly, we plot (a) the map of the intensity of the static electron quasiparticle scattering rate (or so-called the electron quasiparticle elastic-scattering rate) $\Gamma({\bf k},0)$ and (b) the angular dependence of $\Gamma({\bf k}_{\rm F},0)$ along EFS from the antinode to the node at the electron doping $\delta=0.15$ with $T=0.002J$ for $t/J=-2.5$ and $t'/t=0.4$ in Fig. \ref{scattering-rate}. For comparison, the corresponding ARPES experimental result of the electron quasiparticle elastic-scattering rate along EFS observed from the electron-doped cuprate superconductor \cite{Horio16} Pr$_{1.3-x}$La$_{0.7}$Ce$_{x}$CuO$_{4}$ is also presented in Fig. \ref{scattering-rate}b (inset). Moreover, we have also calculated the angular dependence of $\Gamma({\bf k}_{\rm BS},0)$ along the back side of the Fermi pocket from the antinode to the node, and found that its behavior is very similar to that of $\Gamma({\bf k}_{\rm F},0)$. In Fig. \ref{scattering-rate}, it is obvious that the special structure of the experimentally observed electron quasiparticle elastic-scattering rate $\Gamma({\bf k}_{\rm F},0)$ along EFS \cite{Horio16,Santander-Syro09} is qualitatively reproduced, where $\Gamma({\bf k}_{\rm F}, 0)$ has a strong angular dependence with the strongest scattering appeared at the antinodes. It is interesting to note that the strongest electron quasiparticle scattering appeared at the antinodes has been also widely observed in the hole-doped counterparts \cite{Comin14,Valla00,Kaminski05,Shi08,Sassa11}, indicating a common electron quasiparticle scattering mechanism both in the hole- and electron-doped cuprate superconductors. On the other hand, the weakest electron quasiparticle scattering does not take place at the nodes, but occurs exactly at the hot spots ${\bf k}_{\rm HS}$, where the energy spectra $\varepsilon_{0{\bf k}_{\rm HS}}\approx-\varepsilon_{{\bf k}_{\rm HS}}$, and then $E^{+}_{{\bf k}_{\rm HS}}\approx E^{-}_{{\bf k}_{\rm HS}}\approx \varepsilon_{{\bf k}_{\rm HS} }$ for the electron quasiparticle excitation spectra at the antibonding and bonding bands. In particular, the hot spot on EFS in Fig. \ref{scattering-rate} appears at $27^{\rm o}$ of the EFS angle at the electron doping $\delta=0.15$, in good agreement with the corresponding experimental result of $25^{\rm o}$ of the EFS angle \cite{Horio16}. This highly anisotropic momentum dependence of the electron quasiparticle elastic-scattering rates (then the charge-order gap) on ${\bf k}_{\rm F}$ and ${\bf k}_{\rm BS}$ therefore suppresses heavily the spectral weight of the electron quasiparticle excitations on the contours ${\bf k}_{\rm F}$ and ${\bf k}_{\rm BS}$ in the antinodal region, and reduces modestly the spectral weight in the nodal region. This is also why the tips of these disconnected segments on ${\bf k}_{\rm F}$ and ${\bf k}_{\rm BS}$ assemble on the hot spots to form a Fermi pocket around the nodal region, leading to a coexistence of the Fermi arcs and Fermi pockets. At the same time, this EFS instability therefore drives the charge-order correlation with the charge-order wave vector connecting the parallel hot spots on EFS, as the charge-order correlation driven by the EFS instability in the hole-doped counterparts \cite{Lee14,Harrison14,Sachdev13,Meier13,Atkinson15,Feng16}. In other words, the coexistence of the Fermi arcs and Fermi pockets and the charge-order correlation are a natural consequence of the highly anisotropic momentum dependence of the charge-order gap originated from the electron self-energy due to the interaction between electrons by the exchange of spin excitations. However, despite the common physical origin of the EFS instability driven charge-order correlation both in the hole- and electron-doped cuprate superconductors, the position of the hot spots in electron-doped side shifts to the nodes with the increase of the electron doping, which is in a clear contrast to the hole-doped counterparts, where the position of the hot spots moves towards to the antinodes when the hole doping is increased. This subtle difference therefore leads to that the charge-order wave vector in hole-doped case {\it decreases} with the increase of the hole doping \cite{Comin15,Comin14,Feng16}, while it {\it increases} with the increase of the electron doping in the electron-doped side \cite{Neto16}.

\subsection{Quantitative link between single-electron fermiology and collective response of electron density}\label{charge-order}

In order to see if the electron quasiparticle excitations around the hot spots determined by the low-energy electronic structure is quantitatively linked to the charge-order correlation, we now discuss the collective response of the electron density. The dynamical charge structure factor $C({\bf k}, \omega)$ can be obtained directly from the imaginary part of the electron density-density correlation function as,
\begin{eqnarray}\label{DCSF}
C({\bf k},\omega)=-{1\over\omega}{\rm Im}\tilde{\Pi}_{\bf c}({\bf k},\omega),
\end{eqnarray}
where the electron density-density correlation function $\tilde{\Pi}_{\bf c}({\bf k},\omega)$ is defined as,
\begin{eqnarray}\label{EDDCF}
\tilde{\Pi}_{\bf c}({\bf R}_{l}-{\bf R}_{l'},t-t')=\langle\langle T\rho({\bf R}_{l},t)\rho({\bf R}_{l'},t')\rangle\rangle,
\end{eqnarray}
with the electron density operator,
\begin{eqnarray}\label{EDO}
\rho({\bf R}_{l})=e\sum_{\sigma}C^{\dagger}_{l\sigma}C_{l\sigma}.
\end{eqnarray}
Substituting the electron density operator (\ref{EDO}) into Eqs. (\ref{EDDCF}) and (\ref{DCSF}), the dynamical charge structure factor $C({\bf k}, \omega)$ therefore can be obtained explicitly in terms of the electron spectral function (\ref{SF}) as,
\begin{eqnarray}\label{DCSF1}
C({\bf k},\omega)&=&2e^{2}{1\over N}\sum_{\bf q}\int^{\infty}_{-\infty}{d\omega'\over 2\pi}A({\bf q}+{\bf k},\omega'+\omega)A({\bf q},\omega') \nonumber\\
&\times& {n_{\rm F}(\omega'+\omega)-n_{\rm F}(\omega')\over\omega}. ~~~~~~~
\end{eqnarray}
It should be emphasized that in the obtaining above dynamical charge structure factor, the vertex correction has been ignored, since it has been shown that the vertex correction is negligibly small in the calculation of the  density-density correlation of cuprate superconductors \cite{Lin12,Jing17}.

\begin{figure}[h!]
\centering
\includegraphics[scale=1.2]{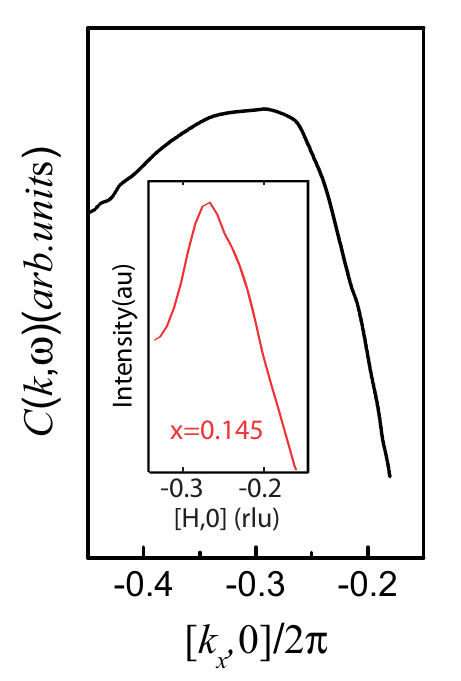}
\caption{(Color online) Dynamical charge structure factor along the ${\bf k}= [0,0]$ to ${\bf k}=[\pi,0]$ direction of the Brillouin zone at $\omega=8.1J$ and $\delta=0.15$ with $T=0.002J$ for $t/J=-2.5$ and $t'/t=0.4$. Inset: the corresponding experimental data of Nd$_{2-x}$Ce$_{x}$CuO$_{4}$ taken from Ref. \onlinecite{Neto15}. \label{DCFS}}
\end{figure}

\begin{figure}[h!]
\centering
\includegraphics[scale=1.2]{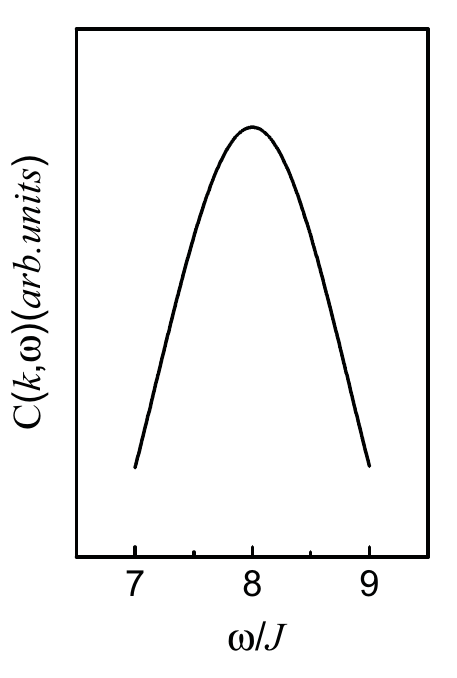}
\caption{Dynamical charge structure factor as a function of energy in the wave vector $Q_{\rm CD}=0.28$ at $\delta=0.15$ with $T=0.002J$ for $t/J=-2.5$ and $t'/t=0.4$. \label{DCFS-energy}}
\end{figure}

To reveal the global feature of the charge-order correlation, we have mapped the dynamical charge structure factor (\ref{DCSF1}) in the $[k_{x},k_{y}]$ plane, and find that there are four resonance peaks, which appear always along the parallel directions $[\pm Q_{\rm HS},0]$ and $[0,\pm Q_{\rm HS}]$ of the Brillouin zone - remarkably similar in the directions to the resonance peak observed in the hole-doped counterparts \cite{Comin15,Campi15,Comin14,Wu11,Chang12,Ghiringhelli12,Neto14,Comin15a}. To show the parallel resonance peaks around the wave vectoe $Q_{\rm HS}$ clearly, the result of $C({\bf k}, \omega)$ along the ${\bf k}= [0,0]$ to ${\bf k}= [\pi,0]$ direction of the Brillouin zone at the energy $\omega=8.1J$ and the electron doping $\delta=0.15$ with $T=0.002J$ for $t/J=-2.5$ and $t'/t=0.4$ is plotted in Fig. \ref{DCFS} in comparison with the corresponding experimental data \cite{Neto16} obtained from the electron-doped cuprate superconductor Nd$_{2-x}$Ce$_{x}$CuO$_{4}$ around the electron doping $\delta=0.145$ (inset). Obviously, an resonance peak in Fig. \ref{DCFS} appears in the wave vector $Q_{\rm CD}\approx 0.28$, closely matching the experimental value of the charge-order wave vector $Q_{\rm CD}\sim 0.27$ found in the electron-doped cuprate superconductors \cite{Neto16}. To confirm this resonance peak that can be identified as the presence of charge ordering, we plot $C(Q_{\rm CD},\omega)$ as a function of energy in the wave vector $Q_{\rm CD}=0.28$ at the electron doping $\delta=0.15$ with $T=0.002J$ for $t/J=-2.5$ and $t'/t=0.4$ in Fig. \ref{DCFS-energy}, where the energy is tuned away from the resonance, the distinct peak in the wave vector $Q_{\rm CD}=0.28$ is suppressed, and then eventually disappears, also in qualitative agreement with the experimental observation from the RXS measurements \cite{Neto15,Neto16}. This suppression of the resonance peak by tuning energy away from the resonance therefore confirms the presence of the charge-order correlation as the collective response of the electron density in the electron-doped cuprate superconductors.

An explanation of the above obtained results for the collective response of the electron density in the electron-doped cuprate superconductors can be found from the single-electron Green's function (\ref{EGF1}) [then the charge-order gap in Eq. (\ref{PG})], which states a fact that the formalism of the single-electron Green's function (\ref{EGF1}) and the related electron antibonding and bonding energy bands $E^{+}_{\bf k}$ and $E^{-}_{\bf k}$ can be also reproduced qualitatively by a phenomenological Hamiltonian,
\begin{eqnarray}\label{CO-model}
H_{\rm CO}&=&\sum_{{\bf k}\sigma}\varepsilon_{\bf k}C^{\dagger}_{{\bf k}\sigma}C_{{\bf k}\sigma}\nonumber\\
&+&\sum_{{\bf k}\sigma}\bar{\Delta}_{\rm CD}({\bf k}) C^{\dagger}_{{\bf k}+{\bf Q}_{\rm HS} /2\sigma} C_{{\bf k}-{\bf Q}_{\rm HS}/2\sigma}.~~~~
\end{eqnarray}
In particular, this type Hamiltonian has been usually employed to phenomenologically discuss the relationship between the EFS instability and the charge-order correlation in cuprate superconductors \cite{Harrison14,Sachdev13,Meier13,Atkinson15}, where the EFS reconstruction is due to the emergence of the charge-order correlation, and then the original EFS is broken up into the Fermi pockets and their coexistence with the Fermi arcs, in qualitative agreement with the experimental data. This is why our present study based on the t-J model in the fermion-spin representation can give a consistent description of the charge-order correlation in the electron-doped cuprate superconductors. In other words, our present theoretical results also show why the qualitative feature of the charge-order correlation observed from the experiments can be fitted by using a similarly phenomenological Hamiltonian (\ref{CO-model}).

\section{Conclusions}\label{conclusions}

In conclusion, within the framework of the $t$-$J$ model in the charge-spin separation fermion-spin representation, we have studied the microscopic origin of the charge-order correlation in the electron-doped cuprate superconductors. We evaluate the electron self-energy in terms of the full charge-spin recombination, and then employ it to calculate the electron spectral function and the dynamical charge structure factor. In particular, we qualitatively reproduce the experimentally observed electron quasiparticle elastic-scattering rate along EFS, which is highly anisotropic in momentum space, and is intriguingly related to the charge-order gap in the electron's band structure. Although the scattering strength appears to be weakest at the hot spots, the stronger scattering is found in the antinodal region than in the nodal region, which therefore leads to that EFS shrinks down to the disconnected segments located around the nodal region. In particular, the tips of these disconnected segments converge on the hot spots to form the Fermi pockets, generating a coexistence of the Fermi arcs and Fermi pockets. Moreover, this EFS instability drives charge order, with the charge-order wave vector that matches well with the wave vector connecting the hot spots on EFS, as charge order in the hole-doped counterparts. However, in a clear contrast to the hole-doped case, the charge-order wave vector in the electron-doped side smoothly increases with the increase of the electron doping. Our theory also shows that there is indeed a quantitative link between the single-electron fermiology and the collective response of the electron density. Incorporating the present result with that obtained previously in the hole-doped counterparts \cite{Feng16}, it is thus shown that there is a common physical origin for the charge-order correlation both in the hole- and electron-doped cuprate superconductors.

\acknowledgments

The authors would like to thank Dr. Deheng Gao, Dr. Yiqun Liu, Dr. Huaisong Zhao, and Professor Yongjun Wang for helpful discussions. This work was supported by the National Key Research and Development Program of China under Grant No. 2016YFA0300304, and National Natural Science Foundation of China under Grant Nos. 11574032 and 11734002.

\end{document}